\documentclass{IEEEtaes}

\usepackage[dvipsnames]{xcolor}
\usepackage{array,amsthm}
\usepackage{graphicx}
\usepackage{amsmath}
\usepackage{amssymb}
\usepackage{bm}
\usepackage{booktabs}
\usepackage{float}
\usepackage{thmtools}
\usepackage{tikz}
\usetikzlibrary{positioning, arrows.meta}
\usepackage{hyperref}

\jvol{XX}
\jnum{XX}
\jmonth{XXXXX}
\paper{1234567}
\pubyear{2026}
\doiinfo{TAES.2026.Doi Number}

\begin{document}
\title{Optimal Trajectory Generation for Improved Magnetic Navigation}
\author{JUSTIN KANG}
\author{TEDDY HERRERA}
\author{LIRAZ MUDRIK, Member, IEEE}
\author{SEAN KRAGELUND}
\author{ALFONSO SCIACCHITANO}
\author{ISAAC KAMINER}
\affil{Naval Postgraduate School, Monterey, CA, USA}

\receiveddate{Manuscript received XXXXX 00, 0000; revised XXXXX 00, 0000; accepted XXXXX 00, 0000. This work was supported in part by the SMART/NAWCWD Fellowship Program, which funded the first author's Ph.D. studies at the Naval Postgraduate School; the work of Sean Kragelund was supported by NPS CRUSER; and the work of Isaac Kaminer was supported by the Office of Naval Research Science of Autonomy Program under Grant No.\ N0001425GI01545.} 

\corresp{{\itshape (Corresponding author: Liraz Mudrik)}.}

\authoraddress{Authors’ address: The authors are with the Department of Mechanical and Aerospace Engineering,
Naval Postgraduate School, Monterey, CA 93943, USA
(e-mail: \href{mailto:liraz109@gmail.com}{liraz109@gmail.com}).
}

\markboth{KANG ET AL.}{OPTIMAL TRAJECTORY GENERATION FOR IMPROVED MAGNETIC NAVIGATION}
\maketitle


\begin{abstract}
Magnetic navigation has emerged as a promising alternative for navigation in Global Positioning System (GPS)-denied environments, leveraging geomagnetic field maps in conjunction with onboard magnetometer measurements. However, its performance is highly sensitive to trajectory-dependent observability, which limits its practical effectiveness under conventional flight paths. This paper proposes an optimal trajectory design framework for magnetic navigation that maximizes information content along the flight path. The trajectory generation problem is formulated as an optimal control problem that minimizes the posterior Cram\'{e}r--Rao lower bound on the position estimation error, subject to a penalty on path length. The resulting trajectories are non-intuitive and significantly enhance the observability of the navigation system. Simulation results demonstrate that the proposed optimal trajectories yield substantial reductions in estimation error compared to conventional straight-line trajectories, highlighting the critical role of trajectory design in enabling high-accuracy magnetic navigation. These findings suggest that trajectory optimization can substantially improve the viability of magnetic navigation as a robust alternative for aerospace applications in GPS-denied environments.
\end{abstract}

\begin{IEEEkeywords}
GPS-denied navigation, magnetic anomaly navigation, posterior Cram\'{e}r--Rao lower bound (PCRLB), pseudospectral optimal control, trajectory optimization.
\end{IEEEkeywords}


\newpage
\section{INTRODUCTION}

Modern aerial navigation systems rely heavily on the integration of inertial navigation systems (INS) with the Global Positioning System (GPS) to achieve accurate position, velocity, and attitude estimation. However, GPS signals are vulnerable to jamming and spoofing, making them unreliable in contested or denied environments. In such scenarios, navigation performance degrades rapidly due to unbounded INS drift, motivating the need for alternative, independent navigation aids.

Several GPS-denied navigation techniques have been explored, including terrain-aided navigation (TAN) and vision-based navigation. TAN relies on terrain elevation matching using digital terrain maps such as Digital Terrain Elevation Data (DTED)~\cite{Siouris_GNC_Book_2004}. While effective in regions with sufficient terrain variation, its performance degrades over flat terrain such as oceans and deserts, and it requires active sensing that may compromise platform stealth. Vision-based navigation similarly depends on environmental features and is degraded in low-visibility conditions such as fog, clouds, or nighttime operation.

Magnetic navigation, which leverages geomagnetic field maps in conjunction with onboard magnetometer measurements, has recently emerged as a promising alternative due to its passive operation and global availability. Unlike TAN and vision-based methods, magnetic navigation is robust to environmental conditions and does not rely on external emissions. However, its performance is strongly dependent on the spatial variation of the geomagnetic field and the quality of available magnetic maps~\cite{Canciani_Raquet_Airborne_2017,Gupta_LowerBounds_2024}.

Despite recent advances, several challenges limit the practical deployment of magnetic navigation. First, the resolution and accuracy of global geomagnetic maps are often insufficient for high-precision navigation. Typical global models have resolutions on the order of thousands of kilometers, while magnetic anomaly maps provide finer resolution on the order of kilometers~\cite{WMM_Report_2025,EMAG2V3_Article_2017}. While localized high-resolution surveys can significantly improve performance, they are not always feasible for large-scale or operational use. How much map quality costs in achievable accuracy has been quantified through Cram\'{e}r--Rao-type lower bounds~\cite{Tkhorenko_Karshakov_2022,Gupta_LowerBounds_2024,Sengupta_Accuracy_2025}; the present work uses the same bound not to analyze a given flight path but to design one.

Second, magnetic measurements are susceptible to interference from onboard electronics, requiring careful calibration and sensor placement. Flight experiments have demonstrated that high-accuracy magnetic navigation is achievable using high-resolution maps and carefully calibrated sensors, achieving a distance root-mean-square (DRMS) error of approximately 13 meters over a one-hour flight at 300\,m above ground level~\cite{Canciani_Raquet_Airborne_2017}. More recent experiments demonstrated 59-meter DRMS accuracy over a 65-minute F-16 flight at the same altitude, using onboard sensors with online Tolles--Lawson calibration~\cite{Canciani_F16_Test_2022}. Field trials with quantum magnetometers have further validated magnetic navigation accuracy at levels exceeding that of strategic-grade INS~\cite{Muradoglu_QCTRL_2025}.

A third and less explored challenge is the dependence of navigation accuracy on the vehicle trajectory. Regions with low magnetic field gradient variation provide limited information for state estimation, leading to poor observability and increased navigation error. This phenomenon is analogous to terrain-aided navigation, where performance degrades over flat terrain~\cite{Siouris_GNC_Book_2004}. Experimental results have shown that navigation accuracy improves significantly when traversing regions with higher magnetic field variability~\cite{Canciani_Raquet_Airborne_2017,Canciani_F16_Test_2022}, suggesting that trajectory design plays a critical role in magnetic navigation performance.

The problem of optimizing observer trajectories to maximize estimation performance has a rich history in the bearings-only localization literature, where the Fisher information matrix (FIM) and the Cram\'{e}r--Rao lower bound have been used to design information-maximizing paths~\cite{Hammer_FIM_1989,Oshman_FIM_1999,Ponda_FIM_2009}. In the context of magnetic navigation, recent work has explored information-aware path planning using observability metrics, entropy reduction, and real-time uncertainty estimation~\cite{Penumarti_RealTime_2026}. These approaches employ receding-horizon or reactive planning strategies that adjust the trajectory at each time step based on current uncertainty estimates. In contrast, the present work formulates the trajectory design as a global optimal control problem that minimizes the posterior Cram\'{e}r--Rao lower bound (PCRLB)~\cite{tichavsky_posterior_1998} over the entire flight path prior to execution, producing a trajectory that is optimal with respect to the complete magnetic field map.

The contribution of this paper is to formulate magnetic navigation trajectory design as an optimal control problem whose terminal cost is the PCRLB of the position estimation error, so that the entire flight path is optimized against the complete magnetic field map before execution rather than adjusted locally during it. Solving this problem requires a gradient-informed starting point, because the terminal cost is nearly insensitive to the trajectory wherever the field is flat; candidate trajectories are therefore first generated by a gradient-weighted A* search~\cite{Lavalle_Trajectory_Planning_Book_2006,Zhang_Trajectory_Planning_2018}, and the optimal control problem is then solved by pseudospectral collocation~\cite{Ross_DIDO_2015,fahroo_direct_2002}.
A simulation study over two geographic regions shows that trajectory design determines whether magnetic aiding helps at all. The optimized trajectories improve on unaided inertial navigation in every case examined, with only a modest increase in path length. A straight-line path through a low-gradient region, by contrast, can collect too little field information to correct the inertial drift.

The remainder of this paper is organized as follows. Section~\ref{sec:background} presents the background on magnetic field modeling and the posterior Cram\'{e}r--Rao lower bound. Section~\ref{sec:problem} formulates the trajectory optimization problem and describes the numerical solution method. Section~\ref{sec:results} presents the simulation results and performance evaluation. Section~\ref{sec:conclusion} concludes the paper. The Appendix collects the navigation filter equations, the relation between the discrete-time information recursion and its continuous-time limit, and a demonstration of Pareto front based weight selection.

\section{BACKGROUND}
\label{sec:background}
The following notation conventions are adopted throughout this paper. Boldface lowercase and uppercase letters denote vectors and matrices, respectively, while scalar quantities appear in standard italic type. The system dynamics are formulated in a local North--East--Down reference frame treated as inertial, and SI units are used unless otherwise stated.
\subsection{Geomagnetic Field Model}
The geomagnetic field can be decomposed into contributions from four primary sources: the core field, the crustal anomaly field, external fields due to solar wind interactions, and local disturbances~\cite{Canciani_Raquet_Airborne_2017,WMM_Report_2025}. The core field, generated by convective motion in the Earth's outer core, dominates the total field with magnitudes ranging from 25,000 to 65,000\,nT and exhibits slow secular variation on a yearly timescale. The crustal anomaly field arises from magnetized rocks in the lithosphere, with magnitudes up to several thousand nT, and is effectively time-invariant for navigation purposes. Solar wind effects introduce hourly variations on the order of tens to hundreds of nT, while local disturbances from onboard electronics can range from $\pm 20$\,nT for isolated sensors to $\pm 10{,}000$\,nT for sensors mounted inside an aircraft~\cite{Canciani_Raquet_Airborne_2017,Canciani_F16_Test_2022}.

For aerial magnetic navigation, the crustal anomaly field is the primary signal of interest, as its spatial variability provides the information content necessary for position estimation. Two properties of the anomaly field are particularly relevant to this work. First, the spatial gradients of the anomaly field decrease with increasing altitude, leading to reduced observability at higher flight levels. Second, the effective spatial resolution of the anomaly field at a given altitude is approximately equal to that altitude, a consequence of the physics governing crustal magnetization~\cite{Canciani_Raquet_Airborne_2017}.

\begin{figure}[t]
\centerline{\includegraphics[width=18pc]{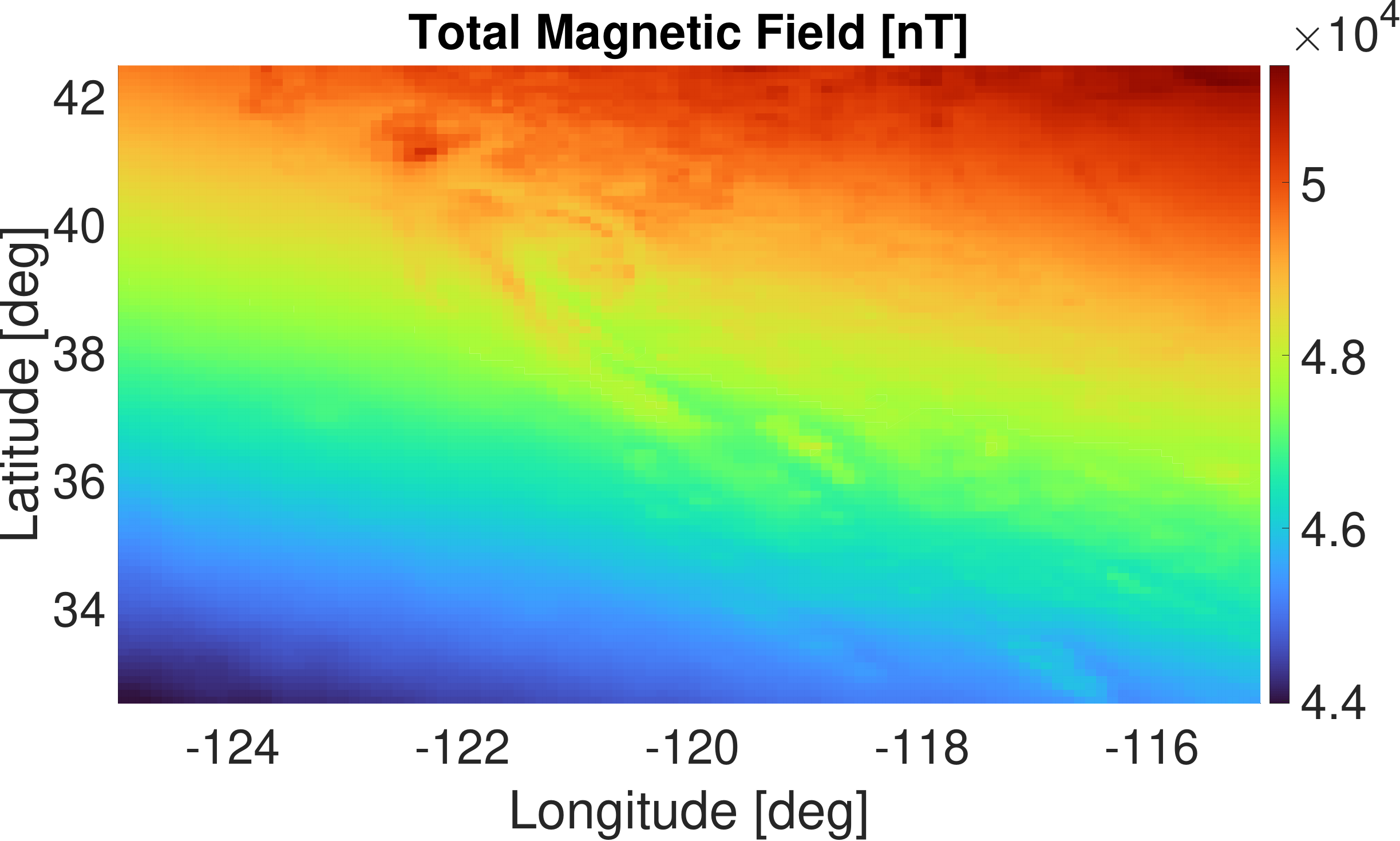}}
\caption{Total magnetic field intensity used in this work, constructed as the sum of the WMM core field and the EMAG2V3 crustal anomaly field~\cite{EMAG2V3_Article_2017,WMM_Report_2025}.}
\label{fig:mag_field_map}
\end{figure}

In this work, the total geomagnetic field is modeled as the sum of the core field and the crustal anomaly field, as illustrated in Fig.~\ref{fig:mag_field_map}. The core field is represented by the World Magnetic Model (WMM), a spherical harmonic expansion of degree and order 12 with a minimum resolved wavelength of approximately 3,200\,km, published by the National Oceanic and Atmospheric Administration~\cite{WMM_Report_2025}. An alternative core field model, the International Geomagnetic Reference Field (IGRF), is maintained by the International Association of Geomagnetism and Aeronomy and uses a spherical harmonic expansion of degree and order 13~\cite{IGRF_13_2021}. The anomaly field is obtained from the Earth Magnetic Anomaly Grid version 3 (EMAG2V3), which provides scalar anomaly values at 2 arc-minute resolution, corresponding to approximately 3.7\,km at the equator~\cite{EMAG2V3_Article_2017}. The remaining field components, including solar wind effects and local disturbances, are treated as additive stochastic noise.
\subsection{Posterior Cram\'{e}r--Rao Lower Bound}

Magnetic navigation accuracy depends on the spatial variation of the magnetic field along the vehicle trajectory. Regions with weak spatial gradients provide limited information for state estimation, resulting in increased position uncertainty. Trajectory design can therefore be used to improve estimation performance by steering the vehicle through regions of high spatial variability. The PCRLB provides the theoretical foundation for quantifying this relationship between trajectory geometry and estimation accuracy.

Consider the discrete-time nonlinear filtering problem defined by
\begin{equation}
\bm{x}_{k+1} = \bm{f}_k(\bm{x}_k) + \bm{v}_k, \label{eq:dynamics_general}
\end{equation}
\begin{equation}
\bm{z}_k = \bm{h}_k(\bm{x}_k) + \bm{w}_k, \label{eq:measurement_general}
\end{equation}
where $\bm{x}_k \in \mathbb{R}^n$ is the state vector at time step $k$, $\bm{z}_k \in \mathbb{R}^m$ is the measurement vector, $\bm{v}_k \sim \mathcal{N}(\bm{0}, \bm{Q}_k)$ is zero-mean Gaussian process noise with covariance $\bm{Q}_k$, and $\bm{w}_k \sim \mathcal{N}(\bm{0}, \bm{R}_k)$ is zero-mean Gaussian measurement noise with covariance $\bm{R}_k$. Under the usual regularity conditions on the joint density, the PCRLB lower-bounds the mean-square estimation error of any Bayesian estimator $\hat{\bm{x}}_k$~\cite{tichavsky_posterior_1998}:
\begin{equation}
\mathrm{E}\!\left[(\hat{\bm{x}}_k - \bm{x}_k)(\hat{\bm{x}}_k - \bm{x}_k)^{T}\right] \succeq \bm{J}_k^{-1},
\label{eq:pcrlb_inequality}
\end{equation}
where $\bm{J}_k$ denotes the Fisher information matrix (FIM) and the inequality is in the positive semidefinite sense.

A recursive formula for computing $\bm{J}_k$ in the general nonlinear filtering setting was derived in~\cite{tichavsky_posterior_1998}. For the system in~\eqref{eq:dynamics_general}--\eqref{eq:measurement_general} with additive Gaussian noise, the recursion is
\begin{equation}
\bm{J}_{k+1} = \bm{D}^{22}_k - \bm{D}^{21}_k \left(\bm{J}_k + \bm{D}^{11}_k\right)^{-1} \bm{D}^{12}_k,
\label{eq:fim_recursion_general}
\end{equation}
where
\begin{subequations}\label{eq:D_matrices}
\begin{align}
&\bm{D}^{11}_k = \bm{F}_k^{T}\, \bm{Q}_k^{-1}\, \bm{F}_k, \label{eq:D11}\\
&\bm{D}^{12}_k = -\bm{F}_k^{T}\, \bm{Q}_k^{-1} = \left(\bm{D}^{21}_k\right)^{T}, \label{eq:D12}\\
&\bm{D}^{22}_k = \bm{Q}_k^{-1} + \bm{H}_{k+1}^{T}\, \bm{R}_{k+1}^{-1}\, \bm{H}_{k+1}, \label{eq:D22}
\end{align}
\end{subequations}
with $\bm{F}_k = \nabla_{\bm{x}} \bm{f}_k\big|_{\bm{x}_k}$ denoting the dynamics Jacobian and $\bm{H}_k = \nabla_{\bm{x}} \bm{h}_k\big|_{\bm{x}_k}$ denoting the measurement Jacobian. The recursion is initialized with $\bm{J}_0 = \bm{P}_0^{-1}$, where $\bm{P}_0$ is the prior state error covariance.

For map-matching navigation systems, it was shown in~\cite{bergman_point-mass_1997} that the PCRLB reduces to the Kalman filter covariance equations with the measurement Jacobian evaluated at the true state. This result, originally derived for terrain-aided navigation, extends directly to magnetic anomaly navigation~\cite{Canciani_Raquet_Airborne_2017}. In the magnetic navigation context, the measurement function $h$ maps the vehicle position to the scalar geomagnetic field intensity at that position, and the measurement Jacobian takes the form
\begin{equation}
\bm{H}_k = \left(\nabla_{\bm{x}} B_m(\bm{x}_k)\right)^{T}\big|_{\bm{x}_k = \bm{x}_k^{\mathrm{true}}},
\label{eq:H_gradient}
\end{equation}
where $B_m$ denotes the total magnetic field intensity and the gradient is taken with respect to the horizontal position coordinates. This result establishes a direct connection between the magnetic field gradient structure and the information available for state estimation: regions with larger spatial gradients yield a more informative measurement Jacobian, which increases the FIM and tightens the PCRLB.

Several scalar metrics can be used to summarize the matrix-valued PCRLB for trajectory optimization, including the determinant, minimum eigenvalue, and trace of the inverse FIM. In this work, the PCRLB trace is adopted as the performance index. It equals the sum of the per-coordinate lower bounds on the mean-square error and therefore bounds the total mean-square position error; because it weights the inverse eigenvalues of the FIM, it is dominated by the least-informed direction, which is the failure mode of interest here~\cite{Ponda_FIM_2009}. This metric is analogous to the position dilution of precision (PDOP) used in GPS navigation, providing an intuitive geometric interpretation of how trajectory geometry affects estimation accuracy~\cite{Ponda_FIM_2009}.

The dependence of the FIM on the trajectory, through the position-dependent measurement Jacobian in~\eqref{eq:H_gradient}, implies that minimizing the PCRLB drives the trajectory toward regions where the field gradient is both large and varying in direction. Each scalar measurement contributes a term proportional to the rank-one matrix $\nabla B_m \nabla B_m^{T}$, which is informative only along the local gradient direction; along a path on which that direction is nearly constant, information accumulates in one direction alone and the trace of the inverse FIM remains large however steep the field. This is the magnetic counterpart of the observer-maneuver requirement in bearings-only localization~\cite{Hammer_FIM_1989,Oshman_FIM_1999}, and it motivates the trajectory optimization framework developed in the following section.


\section{PROBLEM FORMULATION}
\label{sec:problem}

This section formulates the trajectory optimization problem for magnetic navigation. The vehicle dynamics and magnetic field measurement model are introduced first, followed by the cost functional that combines path length and estimation performance as quantified by the PCRLB. The section concludes with a description of the numerical solution method and the navigation filter used for performance evaluation.

\subsection{System Model}

The vehicle is modeled as a point mass operating at constant altitude and constant airspeed in a local North--East--Down frame treated as inertial. Under these assumptions, the motion reduces to a planar problem with heading as the sole degree of freedom. The continuous-time equations of motion are
\begin{subequations}\label{eq:dynamics}
\begin{align}
\dot{x}(t) &= V \cos\theta(t), \label{eq:dyn_x} \\
\dot{y}(t) &= V \sin\theta(t), \label{eq:dyn_y} \\
\dot{\theta}(t) &= u(t), \label{eq:dyn_theta}
\end{align}
\end{subequations}
where $x(t)$ and $y(t)$ are the horizontal position coordinates, $\theta(t)$ is the heading angle, $V$ is the constant airspeed, and $u(t)$ is the heading rate, which serves as the control input. In compact form, the state vector $\bm{x}(t) = [x(t),\, y(t),\, \theta(t)]^{T}$ evolves according to $\dot{\bm{x}}(t) = \bm{f}(\bm{x}(t), u(t))$. The vehicle model represents a long-range missile operating at a constant cruise velocity.

Equations~\eqref{eq:dyn_x}--\eqref{eq:dyn_theta} are the simplified model used for trajectory design. They follow from assuming that the altitude-hold and airspeed-hold autopilot loops maintain level flight at a fixed cruise speed, which reduces the six-degree-of-freedom dynamics to planar kinematics with heading as the sole degree of freedom. Navigation performance is subsequently evaluated on a six-state model that retains altitude, vertical rate, and speed.

The vehicle is equipped with a scalar magnetometer that measures the total geomagnetic field intensity. The measurement at discrete time step $k$ is modeled as
\begin{equation}
z_k = B_m(x_k, y_k) + w_k, \label{eq:mag_measurement}
\end{equation}
where $B_m(x_k, y_k)$ is the total magnetic field intensity at the vehicle position, computed as the sum of the WMM core field and the EMAG2V3 anomaly field as described in Sec.~\ref{sec:background}, and $w_k \sim \mathcal{N}(0, \sigma_{B_m}^2)$ is zero-mean Gaussian measurement noise with variance $\sigma_{B_m}^2$. The pairing of continuous-time dynamics with discrete-time measurements in~\eqref{eq:dynamics} and~\eqref{eq:mag_measurement} is the standard continuous--discrete estimation model~\cite{Taylor1979,Kerr1989}.

\subsection{Trajectory Optimization Problem}

The objective is to find a trajectory from a specified start point to a specified end point that minimizes a weighted combination of path length and position estimation uncertainty. Longer trajectories consume more time and energy but may traverse regions of higher magnetic field gradient, thereby reducing estimation error. The trajectory optimization must balance these competing objectives.

\subsubsection{PCRLB for the Magnetic Navigation Problem}

The PCRLB recursion, presented in Sec.~\ref{sec:background}, is now specialized to the system model introduced above. For trajectory optimization the dynamics are taken to be noise-free, $\bm{Q}_k = \bm{0}$, since the trajectory is the design variable and is prescribed by the optimizer. The estimation uncertainty therefore arises from the initial state uncertainty and the measurement noise alone. The FIM propagation then follows from the standard information filter: at each time step, the prior information is propagated through the state transition and augmented by the new measurement information~\cite{tichavsky_posterior_1998}. This yields the recursion
\begin{equation}
\bm{J}_{k+1} = \bm{\Phi}_{k+1}^{-T}\, \bm{J}_k\, \bm{\Phi}_{k+1}^{-1} + \bm{H}_{k+1}^{T}\, R_{B}^{-1}\, \bm{H}_{k+1},
\label{eq:fim_recursion_specific}
\end{equation}
where $\bm{J}_k$ is the FIM at time step $k$, $\bm{\Phi}_k$ is the discrete-time state transition matrix obtained by discretizing the linearized dynamics, $\bm{H}_k$ is the measurement Jacobian, and $R_B = \sigma_{B_m}^2$ is the scalar magnetometer noise variance. The initial condition is $\bm{J}_0 = \bm{P}_0^{-1}$, where $\bm{P}_0$ is the prior state error covariance, given in Sec.~\ref{sec:sim_setup}. State transition matrices are invertible, so the congruence $\bm{\Phi}_{k+1}^{-T} \bm{J}_k \bm{\Phi}_{k+1}^{-1}$ preserves positive definiteness, and the measurement term in~\eqref{eq:fim_recursion_specific} is positive semidefinite. With $\bm{J}_0 \succ 0$, it follows by induction that $\bm{J}_k \succ 0$ for every $k$ along every candidate trajectory; this uses the absence of process noise and does not carry over to the general recursion~\eqref{eq:fim_recursion_general}.

The dynamics Jacobian, evaluated along the nominal trajectory, is
\begin{equation}
\bm{F} = \left.\frac{\partial \bm{f}}{\partial \bm{x}}\right|_{\bm{x}_{\mathrm{nom}}} =
\begin{bmatrix}
0 & 0 & -V\sin\theta\\
0 & 0 &  V\cos\theta\\
0 & 0 & 0
\end{bmatrix}\bigg|_{\bm{x}_{\mathrm{nom}}},
\label{eq:F_jacobian}
\end{equation}
and the measurement Jacobian is obtained from the gradient of the total magnetic field with respect to position:
\begin{equation}
\bm{H}_k = \left.\left(\nabla_{\bm{x}} B_m\right)^{T}\right|_{\bm{x}_k} = \left[\frac{\partial B_m}{\partial x}\bigg|_{\bm{x}_k},\; \frac{\partial B_m}{\partial y}\bigg|_{\bm{x}_k},\; 0\right],
\label{eq:H_measurement}
\end{equation}
where the partial derivatives are evaluated at the vehicle position $\bm{x}_k$ along the trajectory. The zero in the third component reflects the fact that the scalar magnetic field measurement does not depend on heading. This equation makes explicit the dependence of the FIM on the trajectory: the information accumulated at each time step is determined by the magnetic field gradient at the vehicle position.

\subsubsection{Optimal Control Problem}

The trajectory optimization is formulated as an optimal control problem with continuous-time dynamics and a terminal cost that accumulates at the discrete measurement instants $t_1, \ldots, t_N$. The cost functional is
\begin{equation}
\mathcal{J}(\bm{x}, u) = w_1\, V (t_f - t_0) + w_2\, \mathrm{tr}\!\left(\bm{J}_N^{-1}\right),
\label{eq:cost_functional}
\end{equation}
where $\bm{J}_N$ is the FIM at the final measurement time step and $w_1$, $w_2$ are positive weighting parameters. The first term is the path length traveled: the speed equals the constant $V$, so the length of the path over $[t_0, t_f]$ is $V(t_f - t_0)$, and minimizing it is equivalent to minimizing the flight time. The second term penalizes the PCRLB trace at the terminal time. Because $\bm{J}_N \succ 0$ along every candidate trajectory, the terminal cost is finite and differentiable even for trajectories that traverse regions of weak magnetic gradient; along such segments the bound remains at the level set by the prior, and the cost favors trajectories that add measurement information. Appendix~\ref{app:fim} relates the discrete-time bound $\mathrm{tr}(\bm{J}_N^{-1})$ to its continuous-time counterpart.

The two cost terms carry different units and differ by several orders of magnitude: the PCRLB trace ranges from $10^{-2}$ to a few m$^2$, while the path length is on the order of $10^5$\,m. For weight selection it is therefore convenient to rewrite~\eqref{eq:cost_functional} in the normalized form
\begin{equation}
\mathcal{J}(\bm{x}, u) = w_P\, S_1\, \mathrm{tr}\!\left(\bm{J}_N^{-1}\right) + (1 - w_P)\, S_2\, V (t_f - t_0),
\label{eq:cost_pareto}
\end{equation}
where the scale factors $S_1$ (units of m$^{-2}$) and $S_2$ (units of m$^{-1}$) render both terms dimensionless and of comparable magnitude~\cite{Ross2018}, and the single weight $w_P \in [0, 1]$ sets the trade-off between estimation accuracy and path length. The forms~\eqref{eq:cost_functional} and~\eqref{eq:cost_pareto} are equivalent under $w_1 = (1 - w_P) S_2$ and $w_2 = w_P S_1$. The normalized form has the advantage that sweeping the scalar $w_P$ over $[0, 1]$ traces the attainable trade-offs between the two objectives and supports a Pareto front based selection of the weights~\cite{Hu2013}; the procedure is demonstrated in Appendix~\ref{app:pareto}.

The cost functional~\eqref{eq:cost_functional} is minimized subject to the dynamics~\eqref{eq:dyn_x}--\eqref{eq:dyn_theta}, the initial condition
\begin{equation}
\bm{x}(t_0) = [x_0,\, y_0,\, \theta_0]^{T}, \label{eq:bc_initial}
\end{equation}
the terminal position constraint
\begin{equation}
[x(t_f),\, y(t_f)]^{T} = [x_f,\, y_f]^{T}, \label{eq:bc_terminal}
\end{equation}
and the control constraint
\begin{equation}
|u(t)| \leq u_{\max}, \quad \forall\, t \in [t_0, t_f].
\label{eq:control_constraint}
\end{equation}
The terminal heading $\theta(t_f)$ is left free, and the final time $t_f$ is a free optimization variable. The FIM is computed recursively along the trajectory via~\eqref{eq:fim_recursion_specific}--\eqref{eq:H_measurement} at the discrete measurement times $t_1, t_2, \ldots, t_N$ that lie within the interval $[t_0, t_f]$.

\subsection{Numerical Solution Method}

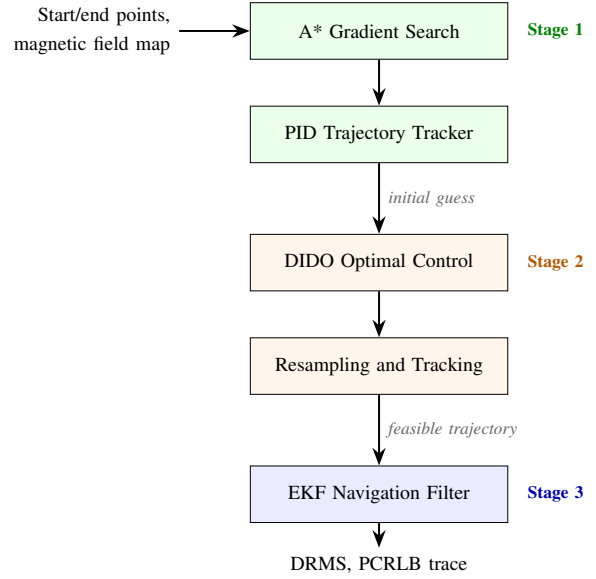
\begin{figure}[ht]
\centering
\resizebox{0.92\columnwidth}{!}{%
\begin{tikzpicture}[
  node distance=0.65cm,
  block/.style={rectangle, draw, text width=3.4cm, minimum height=0.8cm, align=center, font=\small},
  stage/.style={rectangle, rounded corners, draw=gray, dashed, inner sep=5pt},
  arrow/.style={-{Stealth[length=2.5mm]}, thick},
  annot/.style={font=\scriptsize\itshape, text=black!60}
]
\node[block, fill=green!8] (astar) {A* Gradient Search};
\node[block, fill=green!8, below=of astar] (tracker1) {PID Trajectory Tracker};

\node[block, fill=orange!8, below=1.0cm of tracker1] (dido) {DIDO Optimal Control};
\node[block, fill=orange!8, below=of dido] (tracker2) {Resampling and Tracking};

\node[block, fill=blue!8, below=1.0cm of tracker2] (ekf) {EKF Navigation Filter};

\node[right=0.15cm of astar, font=\scriptsize\bfseries, text=green!50!black, anchor=west] {Stage 1};
\node[right=0.15cm of dido, font=\scriptsize\bfseries, text=orange!70!black, anchor=west] {Stage 2};
\node[right=0.15cm of ekf, font=\scriptsize\bfseries, text=blue!70!black, anchor=west] {Stage 3};

\draw[arrow] (astar) -- (tracker1);
\draw[arrow] (tracker1) -- node[right, annot] {initial guess} (dido);
\draw[arrow] (dido) -- (tracker2);
\draw[arrow] (tracker2) -- node[right, annot] {feasible trajectory} (ekf);

\node[left=1.0cm of astar, font=\small, align=right] (inp) {Start/end points,\\magnetic\ field map};
\draw[arrow] (inp) -- (astar);

\node[below=0.35cm of ekf, font=\small, align=center] (out) {DRMS, PCRLB trace};
\draw[arrow] (ekf) -- (out);
\end{tikzpicture}%
}
\caption{Trajectory optimization methodology. Stage~1: gradient-aware initialization via A*. Stage~2: PCRLB-minimizing refinement via optimal control. Stage~3: EKF-based performance evaluation.}
\label{fig:flowchart}
\end{figure}

The problem is non-convex: the dynamics~\eqref{eq:dyn_x}--\eqref{eq:dyn_theta} are nonlinear in the heading, the final time is free, and $\mathrm{tr}(\bm{J}_N^{-1})$ is a non-convex function of the trajectory even for a smooth field. The irregular spatial structure of the magnetic anomaly field adds many local minima. Pseudospectral solvers such as DIDO handle non-convex problems routinely, but this problem presents a further difficulty: wherever the field is flat, $\bm{H}_k \approx \bm{0}$, so the terminal cost barely changes as the trajectory is perturbed and the solver has no descent direction to follow. A straight-line initialization through a low-gradient region therefore stalls where it starts. To address this, a two-stage approach is adopted: a gradient-weighted A* search first places the trajectory where the cost is sensitive to the path, and the optimal control problem is then solved by DIDO. A PID-based trajectory tracker is used between these stages to convert the A* waypoint sequence into a dynamically feasible initial guess for DIDO. 
Finally, trajectory resampling is used to convert 
collocation-based trajectories to uniformly sampled, dynamically consistent paths. The overall methodology is summarized in Fig.~\ref{fig:flowchart}.

\subsubsection{A* Gradient Search for Initialization}
\label{sec:astar}

An initial trajectory is generated using the A* search algorithm~\cite{Lavalle_Trajectory_Planning_Book_2006, Zhang_Trajectory_Planning_2018} with a cost function that balances path length against magnetic field gradient magnitude. The discrete cost function $\mathcal{J}_{\mathrm{AG}}$ for the A* search is
\begin{equation}
\mathcal{J}_{\mathrm{AG}} = \alpha_1\sum_{k=1}^{N-1} \left\|\Delta\bm{r}_{k+1}\right\| - \alpha_2\sum_{k=1}^{N-1} \left\|\nabla B_m(\bm{r}_{k+1})\right\|,
\label{eq:AG_cost}
\end{equation}
where $\bm{r}_k = [x_k,\, y_k]^{T}$ denotes the waypoint position and $\Delta\bm{r}_{k+1} = \bm{r}_{k+1} - \bm{r}_k$ is the position increment between successive waypoints, so the first sum is the discrete path length of the waypoint sequence. The negative sign on the gradient term ensures that the A* minimization favors paths through regions of higher magnetic field variation.

The A* algorithm operates on a discrete grid and does not account for vehicle dynamics, so the resulting A* Gradient (AG) trajectory is generally not dynamically feasible. To produce a flyable path, the AG trajectory is passed through a PID-based trajectory tracker that generates control inputs satisfying the vehicle dynamics~\eqref{eq:dyn_x}--\eqref{eq:dyn_theta}. The tracked AG trajectory then serves as the initial guess for the optimal control solver. However, the AG trajectory optimizes a pointwise proxy for information content (the local gradient magnitude) rather than the actual PCRLB, and does not account for how estimation uncertainty propagates through the state transition over time. The pseudospectral solution in the following subsection addresses these limitations.

\subsubsection{Pseudospectral Optimal Control Solution}

The optimal control problem is solved using DIDO, a pseudospectral optimal control solver based on Legendre--Gauss--Lobatto (LGL) collocation~\cite{Ross_DIDO_2015, fahroo_direct_2002}. DIDO transcribes the continuous-time optimal control problem into a nonlinear programming problem by representing the state and control histories as polynomials interpolated at the LGL nodes. The cost functional~\eqref{eq:cost_functional} and constraints~\eqref{eq:bc_initial}--\eqref{eq:control_constraint} are enforced at the collocation points, and the PCRLB recursion~\eqref{eq:fim_recursion_specific} is evaluated along the discretized trajectory.

The AG trajectory provides a gradient-informed initial guess that accelerates convergence of the nonlinear programming solver. Without this initialization, the solver may converge to a local minimum that does not exploit the magnetic field gradient structure.

For cost evaluation, the DIDO solution must be brought to the uniform sampling grid of the magnetometer: the recursion~\eqref{eq:fim_recursion_specific} presupposes a fixed sampling interval, and evaluating it directly on the non-uniform LGL nodes mis-weights the information contribution of each measurement (Appendix~\ref{app:fim}). The state and control histories $x(t)$, $y(t)$, and $u(t)$ are therefore interpolated in time by the Lagrange polynomials associated with the LGL nodes and evaluated on the uniform grid, using the barycentric form of the interpolant for numerical stability~\cite{Barycentric_Lagrange_Berrut}. Because the interpolant satisfies the dynamics only at the LGL nodes, the resampled path is passed through the trajectory tracker of Sec.~\ref{sec:astar}, which integrates~\eqref{eq:dyn_x}--\eqref{eq:dyn_theta} and therefore returns a trajectory that is dynamically consistent at the uniform time steps. The navigation filter is evaluated on this resampled trajectory.

\subsection{Navigation Performance Evaluation}

The navigation performance of each trajectory is evaluated using an extended Kalman filter (EKF) with a six-state model: $\bm{x}_{\mathrm{EKF}} = [x,\, y,\, z,\, v,\, \dot{z},\, \theta]^{T}$, comprising horizontal position, altitude, speed, vertical rate, and heading. The filter fuses inertial measurements with scalar magnetometer and altimeter measurements to estimate the vehicle state. The six-state model is used for performance evaluation to capture INS drift effects from accelerometer and gyroscope biases, while the three-state model used for trajectory optimization captures the trajectory-dependent observability structure at reduced computational cost. The complete EKF equations are provided in Appendix~\ref{app:ekf}.


\section{RESULTS}
\label{sec:results}

\subsection{Simulation Setup}
\label{sec:sim_setup}

\begin{figure}[ht]
\centerline{\includegraphics[width=19pc]{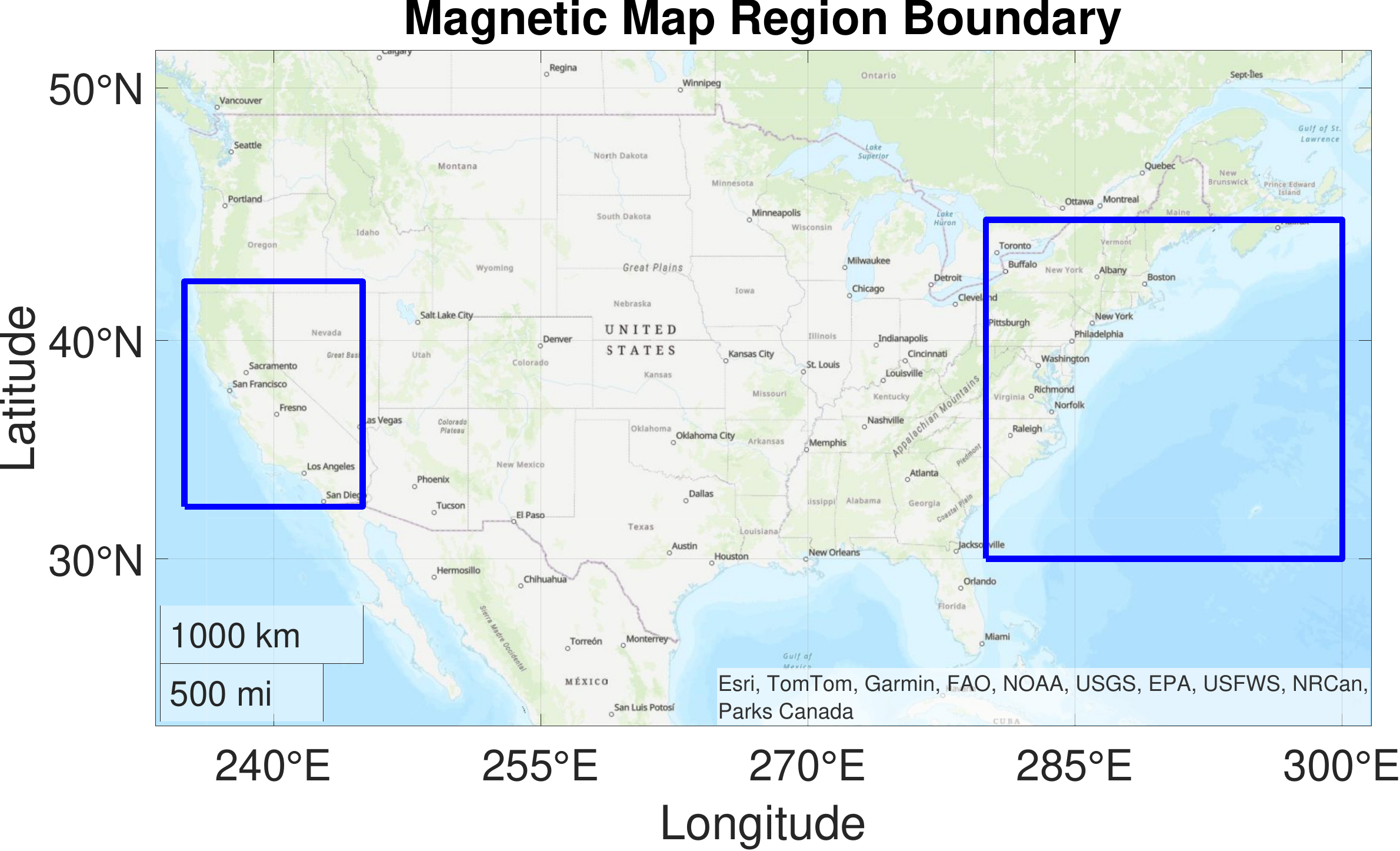}}
\caption{Geographic overview showing the California and New York case study regions.}
\label{fig:us_map}
\end{figure}

\begin{figure}[ht]
\centerline{\includegraphics[width=19pc]{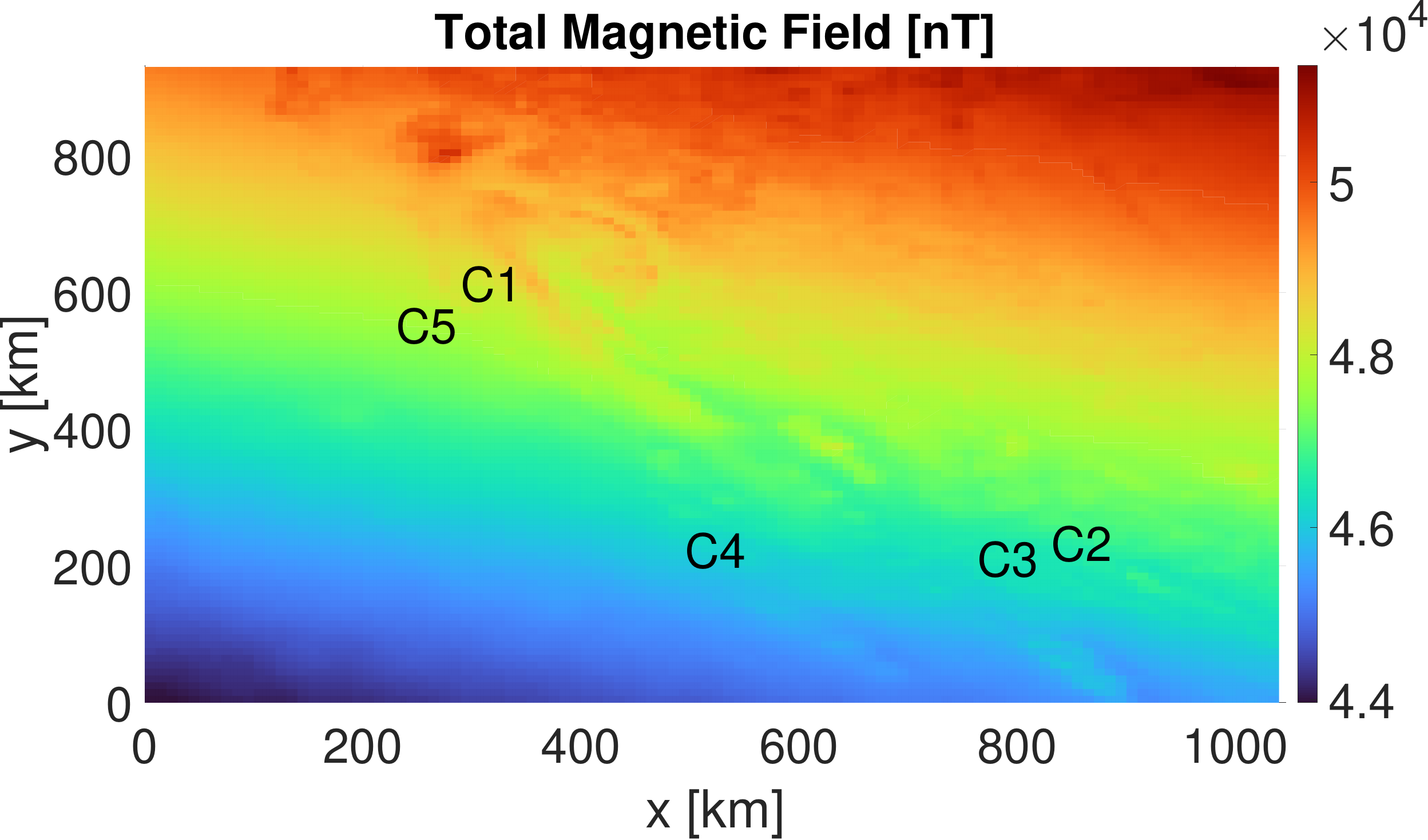}}
\caption{Total magnetic field intensity over the California region with case study locations.}
\label{fig:CA_mag}
\end{figure}

\begin{figure}[ht]
\centerline{\includegraphics[width=19pc]{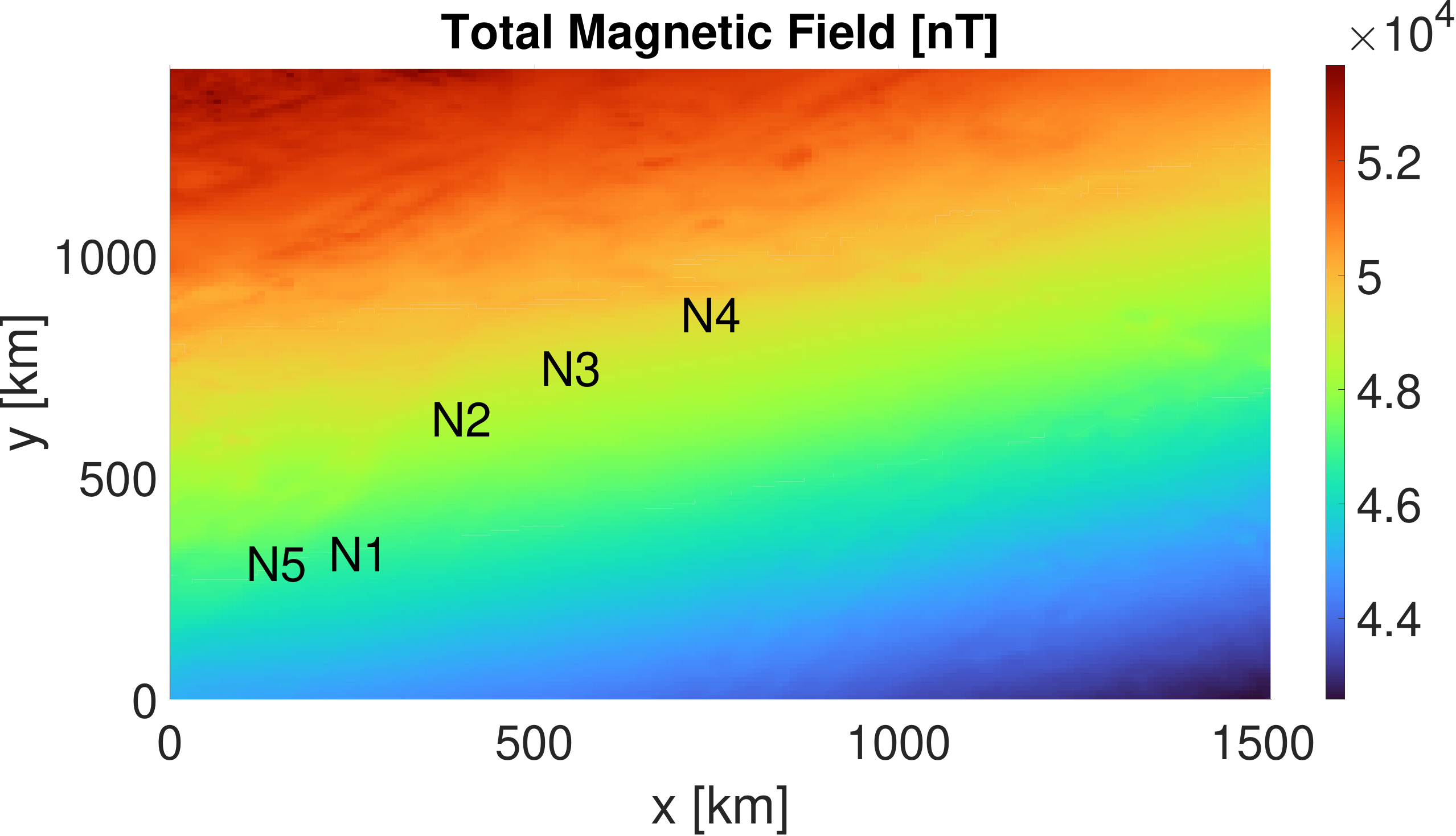}}
\caption{Total magnetic field intensity over the New York region with case study locations.}
\label{fig:NY_mag}
\end{figure}

To evaluate the proposed trajectory optimization framework, case studies were conducted across two geographically distinct coastal regions: a California coastal area and a New York coastal area, as shown in Fig.~\ref{fig:us_map}. Five trajectory pairs were defined in each region, yielding ten case studies in total. Coastal environments were selected to exploit natural contrasts in magnetic field gradients, where ocean regions exhibit relatively low gradients and land regions exhibit higher spatial variability. The total magnetic field intensity over each region is shown in Figs.~\ref{fig:CA_mag} and~\ref{fig:NY_mag}.

For each case study, a straight-line trajectory and a DIDO-FIM optimized trajectory were generated between identical start and end points. The straight-line trajectories were intentionally routed through low-gradient regions, while the DIDO-FIM trajectories were optimized to traverse higher-gradient regions with only modest increases in path length. Navigation performance was evaluated through 1,000 Monte Carlo EKF simulations for each trajectory, with initial conditions and measurements perturbed by Gaussian noise.

The simulation parameters are selected to represent a long-range missile application. The cruise velocity is $V = 224$\,m/s (approximately Mach 0.7). The magnetometer noise standard deviation is $\sigma_{B_m} = 1$\,nT, and the measurement sampling frequency is 1\,Hz. The flight altitude is set to 4\,km, consistent with the EMAG2V3 anomaly grid altitude. The prior state error covariance is $\bm{P}_0 = \bm{I}_{3\times3}$ in SI units, that is 1\,m$^2$ in each position coordinate and 1\,rad$^2$ in heading. The heading prior is left uninformative because the scalar magnetometer measurement carries no direct heading information, as the zero third component of~\eqref{eq:H_measurement} shows. The same $\bm{P}_0$ initializes the navigation filter, so the bound and the filter start from identical prior information. The anomaly map is treated as exact in both the trajectory optimization and the navigation filter. The control bound is set to $u_{\max} = 200$\,deg/s. This is a numerical device rather than a vehicle limit: it is deliberately loose so that the trajectory shape is determined by the cost functional, and it is inactive along all reported solutions, whose turning radii are kilometer-scale (Fig.~\ref{fig:traj_example}). The weighting parameter $w_1 = 0.001$ is held constant across all cases, while $w_2$ is selected manually per case within the range 5 to 500 to balance the path-length and PCRLB-trace contributions to the cost; a systematic alternative to this manual selection is demonstrated in Appendix~\ref{app:pareto}. The number of DIDO collocation points ranges from 10 to 25, selected to balance solution accuracy against polynomial overfitting.

\subsection{Trajectory Characteristics}

\begin{figure}[ht]
\centerline{\includegraphics{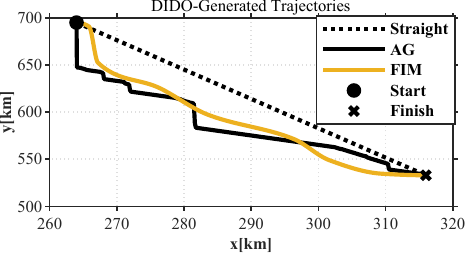}}
\centerline{\includegraphics{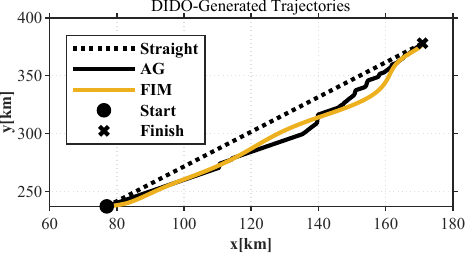}}
\caption{Representative trajectories for Case C1 (Top) and Case N5 (Bottom): straight-line (dotted), AG (black), and DIDO-FIM (yellow).}
\label{fig:traj_example}
\end{figure}

Representative optimized trajectories are shown in Fig.~\ref{fig:traj_example} for one California case (C1) and one New York case (N5). The A* Gradient (AG) trajectories (black) exhibit jagged, non-smooth paths due to the absence of dynamic constraints and serve only as initialization for the DIDO solver. The DIDO-FIM trajectories (yellow) are smooth and dynamically feasible while closely following the gradient-rich structure of the AG paths. Straight-line trajectories are shown as dotted lines for reference.

On average across all ten cases, the DIDO-FIM trajectories deviate from the straight-line paths to traverse regions of higher magnetic field variability, incurring only modest increases in path length (between 1\% and 9\%, 5\% on average, as can be verified from the distance columns in Table~\ref{tab:main_results}). The resulting trajectories are smooth and dynamically feasible, and the small path length penalty is a direct consequence of the PCRLB-based cost functional, which balances information gain against trajectory efficiency.

\subsection{Navigation Performance}

Table~\ref{tab:main_results} presents the complete navigation performance results across all ten cases. For each case, the table reports the trajectory distance, EKF DRMS error from 1,000 Monte Carlo runs, and the PCRLB trace for both the straight-line and DIDO-FIM trajectories.

All five California straight-line cases and three of five New York straight-line cases (N1, N3, N5) diverged. Divergence is defined as the EKF DRMS error exceeding that of the INS-only baseline (approximately 1\,km in most cases), meaning the magnetically-aided filter performs worse than unaided dead-reckoning. In contrast, all ten DIDO-FIM trajectories maintained stable filter performance, with DRMS errors ranging from 88 to 338\,m. A single filter tuning is used throughout, chosen for the information-rich environment encountered along the optimized trajectories and applied unchanged to the straight-line trajectories. The divergences should therefore be read not as a like-for-like accuracy comparison but as a statement about filter design: a magnetic navigation filter tuned for a well-observed flight path over-trusts the magnetic measurements when it is flown through a low-gradient region, and the resulting solution is worse than unaided dead reckoning. Adaptive gain scheduling that accounts for local information content would mitigate this and is left to future work. The accuracy comparison proper is provided by the two cases in which both filters converge, discussed next.

The benefit of trajectory optimization is established by the two cases in which both the straight-line and DIDO-FIM filters converged (N2 and N4), enabling a direct comparison under stable filter operation. For Case N2, the DIDO-FIM trajectory reduces the DRMS error from 251\,m to 151\,m (a 40\% reduction), while the PCRLB trace decreases from 3.76\,m$^2$ to 0.40\,m$^2$. For Case N4, the improvement is even more pronounced: DRMS decreases from 452\,m to 88\,m (81\% reduction), and the PCRLB trace drops from 1.01\,m$^2$ to 0.50\,m$^2$. These non-diverged cases provide the cleanest evidence that the PCRLB-based trajectory optimization produces significant improvements in navigation accuracy independent of the divergence phenomenon.

The PCRLB trace values in Table~\ref{tab:main_results} provide additional insight into the mechanism underlying these results. Across all ten cases, the DIDO-FIM trajectories achieve lower PCRLB traces than the corresponding straight-line trajectories, reflecting the greater position information accumulated along the optimized paths. The magnitude of improvement varies with the geographic characteristics of each case: cases with the largest PCRLB trace reduction (e.g., N1, where the trace decreases from 3.15 to 0.29\,m$^2$) also tend to show the largest DRMS improvement, confirming the theoretical link between the PCRLB and achievable navigation accuracy established in Sec.~\ref{sec:background}.

An important question raised by these results is why Cases N2 and N4 maintain stable straight-line filter performance while the other eight cases diverge. The PCRLB trace alone does not predict this: Case N5 has a lower straight-line PCRLB trace (0.78\,m$^2$) than Case N4 (1.01\,m$^2$), yet N5 diverges while N4 does not. A likely explanation is that the PCRLB trace, being a terminal quantity, does not capture the temporal distribution of information along the trajectory. A path that passes through a brief high-gradient segment surrounded by long low-gradient stretches may achieve a reasonable terminal PCRLB trace, but the filter can drift beyond the linearization region during the low-gradient segments before the informative measurements arrive. The spatial distribution of gradient along the path, not merely its aggregate, appears to determine filter stability.

It is worth noting that the PCRLB provides a lower bound on estimation error and is not expected to predict the absolute EKF performance. For the DIDO-FIM trajectories, the actual DRMS errors (88--338\,m) exceed the square root of the PCRLB trace (0.1--1.0\,m) by two orders of magnitude. This gap is expected, as the PCRLB represents the theoretical minimum achievable by any estimator under the assumed noise model, while the EKF is a suboptimal estimator operating on a nonlinear system with map discretization and interpolation effects. The two quantities are also computed on different models: the bound uses the three-state deterministic model of Sec.~\ref{sec:problem}, whereas the filter uses the six-state model of Appendix~\ref{app:ekf} with process noise, and therefore absorbs INS drift that the bound does not represent. The value of the PCRLB for trajectory optimization lies not in its absolute magnitude but in its ability to rank competing trajectories between the same endpoints: in all ten cases the DIDO-FIM trajectory attains both a lower PCRLB trace and a lower DRMS error than the straight-line trajectory over the same leg. The bound does not rank trajectories across different legs, as the straight-line columns of Table~\ref{tab:main_results} show, since the information available depends on the map in a way that differs from case to case.

The practical cost-benefit trade-off strongly favors the optimized trajectories. As shown in Table~\ref{tab:main_results}, the DIDO-FIM trajectories increase path length by only 1--9\% relative to the straight-line distances. In return, the DRMS improvement ranges from 40\% in the mildest non-diverged case (N2: 251\,m to 151\,m) to over 99\% in the most extreme diverged case (N1: 39,423\,m to 147\,m). Even restricting attention to the two non-diverged cases, where the comparison is fairest, the DIDO-FIM trajectories reduce DRMS by 40--81\% at the cost of 9--13\,km of additional path length. This asymmetry between cost and benefit suggests that PCRLB-based trajectory optimization is a practical tool for mission planning in magnetic navigation applications.

\begin{figure}[ht]
\centerline{\includegraphics[width=18pc]{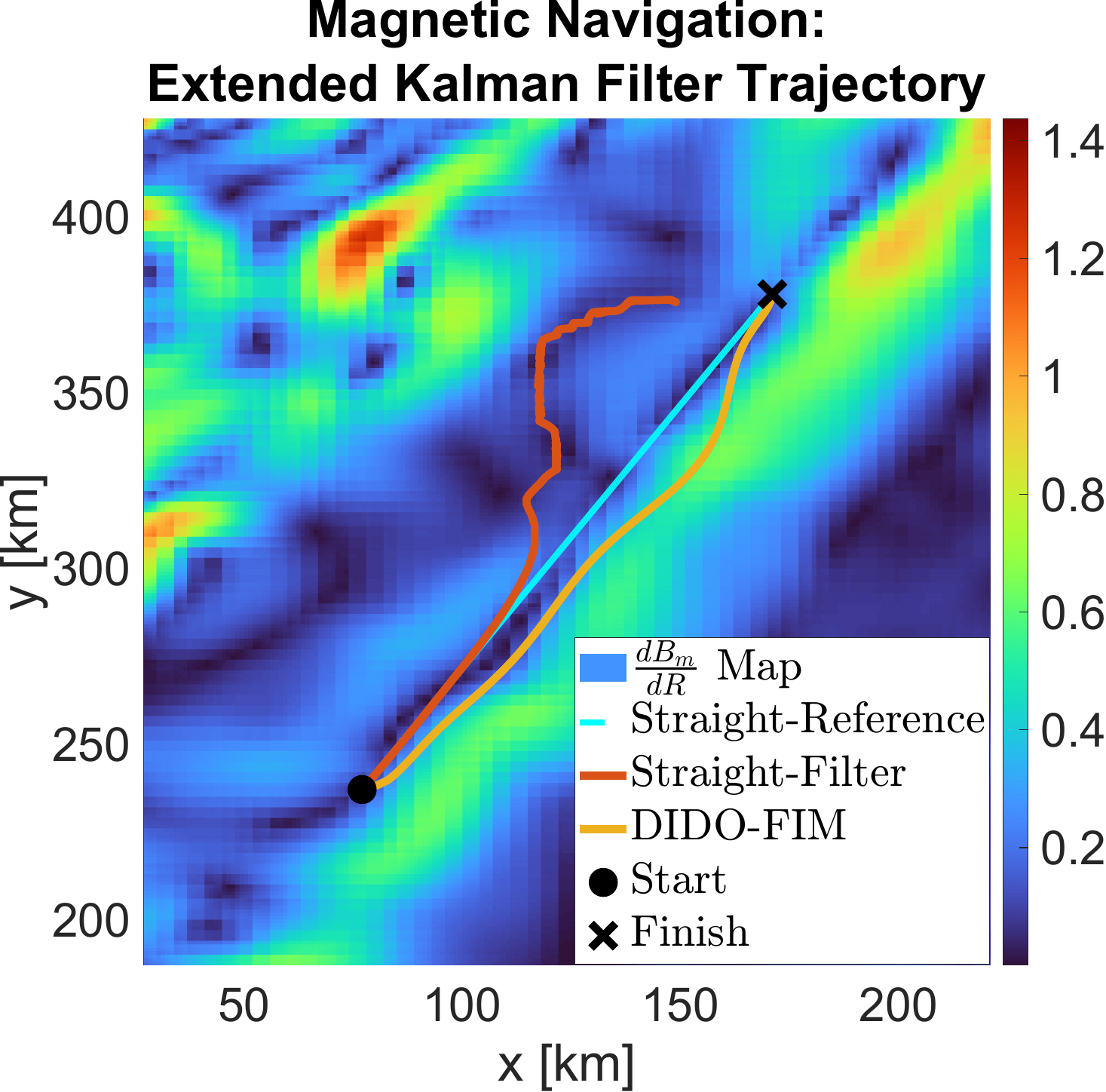}}
\caption{Case N5: EKF trajectory estimates for the straight-line and DIDO-FIM trajectories overlaid on the magnetic field gradient map. The straight-line estimate diverges while the DIDO-FIM estimate tracks the true path.}
\label{fig:Case_N5_Trajectory_3D}
\end{figure}

\begin{figure}[ht]
\centerline{\includegraphics[width=18pc]{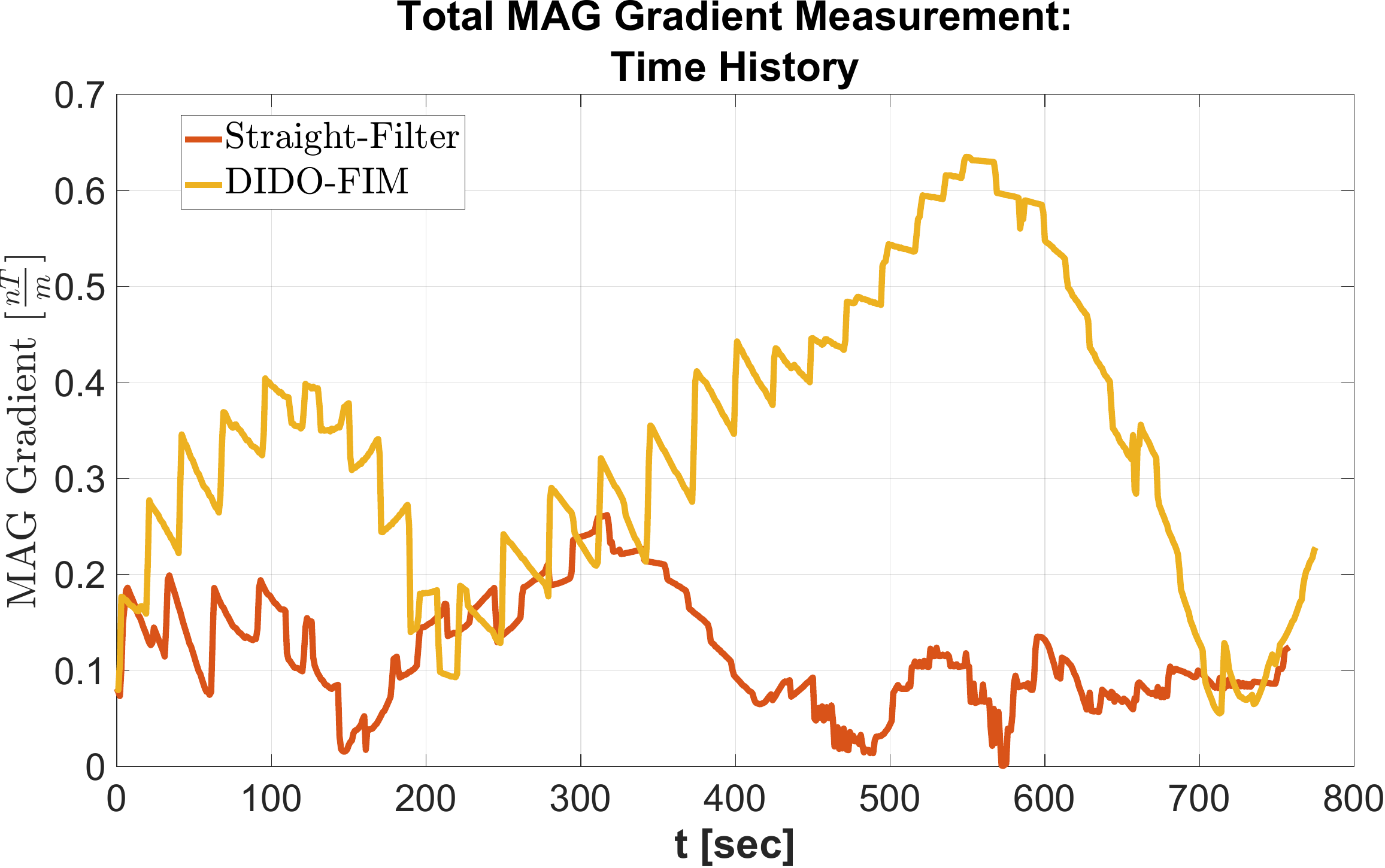}}
\caption{Case N5: Magnetic field gradient magnitude time history showing consistently higher gradients along the DIDO-FIM path.}
\label{fig:Case_N5_Total_Gradient}
\end{figure}

Representative results for Case N5 are presented in Figs.~\ref{fig:Case_N5_Trajectory_3D} and~\ref{fig:Case_N5_Total_Gradient}. Figure~\ref{fig:Case_N5_Trajectory_3D} shows that the straight-line EKF estimate diverges from the true trajectory, while the DIDO-FIM estimate closely tracks the true path throughout the flight. Figure~\ref{fig:Case_N5_Total_Gradient} illustrates the underlying mechanism: the straight-line trajectory passes through regions of low magnetic field gradient, while the DIDO-FIM trajectory traverses regions of consistently higher gradient magnitude.

\begin{table*}[ht]
\centering
\caption{Navigation performance for all ten cases (1,000 Monte Carlo runs per case). The average INS-only DRMS across all cases is approximately 1\,km.}
\label{tab:main_results}
\begin{tabular}{l ccc ccc}
\toprule
 & \multicolumn{3}{c}{Straight-Line} & \multicolumn{3}{c}{DIDO-FIM} \\
\cmidrule(lr){2-4} \cmidrule(lr){5-7}
Case & Dist.\ [km] & DRMS [m] & PCRLB Tr.\ [m$^2$] & Dist.\ [km] & DRMS [m] & PCRLB Tr.\ [m$^2$] \\
\midrule
C1 & 170 & 1538   & 0.020 & 185 & 101 & 0.010 \\
C2 & 336 & 15101  & 0.011 & 339 & 338 & 0.009 \\
C3 & 242 & 30742  & 0.036 & 248 & 142 & 0.032 \\
C4 & 178 & 24525  & 0.51  & 185 & 194 & 0.19  \\
C5 & 178 & 1030   & 1.93  & 192 & 278 & 0.38  \\
\midrule
N1 & 222 & 39423  & 3.15  & 232 & 147 & 0.29  \\
N2 & 181 & 251    & 3.76  & 194 & 151 & 0.40  \\
N3 & 165 & 6405   & 4.72  & 180 & 125 & 0.95  \\
N4 & 164 & 452    & 1.01  & 173 & 88  & 0.50  \\
N5 & 169 & 27549  & 0.78  & 173 & 88  & 0.15  \\
\bottomrule
\end{tabular}
\end{table*}

\section{CONCLUSION}
\label{sec:conclusion}

This paper presented a trajectory optimization framework for magnetic navigation that minimizes the posterior Cram\'{e}r--Rao lower bound (PCRLB) on the position estimation error, and thereby maximizes the information the magnetic field carries about position along the flight path. Through ten case studies across two geographically distinct coastal regions, the optimized DIDO-FIM trajectories were shown to yield substantial improvements in navigation accuracy compared to conventional straight-line trajectories. The optimized trajectories incur only a modest increase in path length (approximately 5\% on average) while reducing the DRMS error to between 88 and 338\,m, against an inertial-only baseline of approximately 1\,km. In eight of the ten cases the straight-line trajectory left the aided estimate worse than unaided dead reckoning, whereas all ten optimized trajectories improved on it; in the two cases where both filters converged, the optimized trajectory reduced the DRMS error by a further 40\% to 81\%.

The results confirm that trajectory design plays a critical role in magnetic navigation by directly influencing estimation observability through the trajectory-dependent measurement Jacobian and the Fisher information matrix. By incorporating the PCRLB into the trajectory optimization cost functional, the proposed approach enables principled selection of flight paths that maximize the information content of the magnetic field measurements. In addition, a normalized form of the cost with a single trade-off weight, swept over its range and filtered for dominance, provided a systematic replacement for the manual weight selection and, on a representative case, reduced both the bound and the realized navigation error at equal path length.


\section*{APPENDIX}
\setcounter{subsection}{0}

\subsection{EKF Implementation}
\label{app:ekf}

The navigation filter uses a six-state extended Kalman filter. The continuous-time system dynamics and measurement model are
\begin{equation}
\dot{\bm{x}} = \bm{f}(\bm{x}, \bm{u}, t) + \bm{v}, \quad \bm{v} \sim \mathcal{N}(\bm{0}, \bm{Q}), \label{eq:ekf_dynamics}
\end{equation}
\begin{equation}
\bm{y} = \bm{h}(\bm{x}) + \bm{w}, \quad \bm{w} \sim \mathcal{N}(\bm{0}, \bm{R}), \label{eq:ekf_measurement}
\end{equation}
where $\bm{x} = [x,\, y,\, z,\, v,\, \dot{z},\, \theta]^{T}$ is the state vector, $\bm{y} = [z_{\mathrm{alt}},\, B_m]^{T}$ is the measurement vector consisting of altitude and total magnetic field, $\bm{v}$ is the process noise, and $\bm{w}$ is the measurement noise. The measurement vector is denoted $\bm{y}$ in this section to distinguish it from the altitude state $z$ contained in $\bm{x}$. The nonlinear dynamics are
\begin{equation}
\bm{f}(\bm{x}, \bm{u}) =
\begin{bmatrix}
v \cos\theta \\ v \sin\theta \\ \dot{z} \\ f_v \\ f_z - g \\ \omega
\end{bmatrix},
\label{eq:ekf_f}
\end{equation}
where $\omega$, $f_v$, and $f_z$ are the heading rate, longitudinal specific force, and vertical specific force, respectively, obtained from the inertial measurement unit (IMU).

For discrete-time EKF implementation, the dynamics are linearized using the Jacobians
\begin{equation}
\bm{F} = \left.\frac{\partial \bm{f}}{\partial \bm{x}}\right|_{\bm{x}_{\mathrm{nom}}}, \quad
\bm{G} = \left.\frac{\partial \bm{f}}{\partial \bm{u}}\right|_{\bm{u}_{\mathrm{nom}}}, \quad
\bm{H} = \left.\frac{\partial \bm{h}}{\partial \bm{x}}\right|_{\bm{x}_{\mathrm{nom}}},
\label{eq:ekf_jacobians}
\end{equation}
and discretized using the state transition matrix $\bm{\Phi}_k = e^{\bm{F}\Delta t}$, the discrete input matrix $\bm{G}_k = \int_0^{\Delta t} e^{\bm{F}\tau} \bm{G}\, d\tau$, and the discrete process noise covariance $\bm{Q}_k$, which is obtained via the Van Loan method or Taylor series expansion. The discrete measurement model uses $\bm{H}$ and $\bm{R}$ unchanged, where the magnetometer entry of $\bm{R}$ is the variance $R_B$ of Sec.~\ref{sec:problem}.

The navigation filter integrates measurements from an IMU, an altimeter, and a scalar magnetometer. The scalar magnetometer measures the total geomagnetic field intensity $\widetilde{B}_T$, from which the interpolated core field $\hat{B}_C$ (based on the estimated position and the WMM) is subtracted to obtain the measured anomaly field $\widetilde{B}_A$. The innovation $\widetilde{B}_A - \hat{B}_A$, where $\hat{B}_A$ is interpolated from the EMAG2V3 map at the estimated position, drives the EKF measurement update.

\subsection{Discrete-Time Information Recursion and Its Continuous-Time Limit}
\label{app:fim}

This appendix records the relation between the discrete-time FIM computed by~\eqref{eq:fim_recursion_specific} and the continuous-time information bound: the discrete matrix is dominated by the continuous one for any sampling interval and converges to it as the interval shrinks, which justifies the use of the discrete recursion as the terminal cost and explains why the recursion must be evaluated on a uniform sampling grid.

The discrete measurement model~\eqref{eq:mag_measurement} samples an underlying continuous measurement record
\begin{equation}
z(t) = B_m(x(t), y(t)) + w(t),
\label{eq:app_cont_meas}
\end{equation}
where $w(t)$ is zero-mean white noise with intensity $R_c$. The discrete measurement is the average of the record over one sampling interval,
\begin{equation}
z_k = \frac{1}{\Delta t}\int_{t_k - \Delta t}^{t_k} z(t)\, dt,
\label{eq:app_avg}
\end{equation}
whose noise variance is
\begin{equation}
R_B = \frac{R_c}{\Delta t}.
\label{eq:app_noise_scaling}
\end{equation}
A shorter interval therefore yields more samples of proportionally lower individual quality, and the information rate of the record is preserved. Equation~\eqref{eq:app_noise_scaling} is the reason the recursion~\eqref{eq:fim_recursion_specific} presupposes a uniform sampling interval: the information contribution of each measurement is weighted by $R_B^{-1} = \Delta t\, R_c^{-1}$, so evaluating the recursion on a non-uniform grid with a fixed $R_B$ mis-weights the contributions.

Let $\bm{\Phi}(t, s)$ denote the state transition matrix of the linearized dynamics $\dot{\bm{x}} = \bm{F}(t)\, \bm{x}$ along the nominal trajectory. For deterministic dynamics, the continuous-time information matrix $\bm{J}(t)$ evolves by the Lyapunov differential equation
\begin{equation}
\dot{\bm{J}} = -\bm{F}^{T} \bm{J} - \bm{J}\, \bm{F} + \bm{H}^{T} R_c^{-1} \bm{H},
\quad
\bm{J}(t_0) = \bm{P}_0^{-1},
\label{eq:app_lyapunov}
\end{equation}
whose solution is
\begin{equation}
\bm{J}(t_f)
=
\bm{\Phi}(t_f, t_0)^{-T} \bm{P}_0^{-1} \bm{\Phi}(t_f, t_0)^{-1}
+
\int_{t_0}^{t_f} \bm{g}(s)\, ds,
\label{eq:app_cont_solution}
\end{equation}
where
\begin{equation}
\bm{g}(s)
=
\bm{\Phi}(t_f, s)^{-T}\, \bm{H}^{T}(s)\, R_c^{-1}\, \bm{H}(s)\, \bm{\Phi}(t_f, s)^{-1}
\succeq 0.
\label{eq:app_integrand}
\end{equation}
Unrolling the recursion~\eqref{eq:fim_recursion_specific} from $\bm{J}_0 = \bm{P}_0^{-1}$ and substituting~\eqref{eq:app_noise_scaling} gives the discrete counterpart
\begin{equation}
\bm{J}_N
=
\bm{\Phi}(t_f, t_0)^{-T} \bm{P}_0^{-1} \bm{\Phi}(t_f, t_0)^{-1}
+
\sum_{k=1}^{N} \bm{g}(t_k)\, \Delta t,
\label{eq:app_disc_solution}
\end{equation}
which contains the same propagated prior and replaces the integral by its right-endpoint Riemann sum.

Two relations between~\eqref{eq:app_cont_solution} and~\eqref{eq:app_disc_solution} follow. First, by~\eqref{eq:app_avg} the samples are a deterministic function of the record, so by the data processing inequality for Fisher information~\cite{Zamir1998} the information they carry cannot exceed that of the record. The ordering is information-theoretic; it does not follow from comparing the Riemann sum in~\eqref{eq:app_disc_solution} with the integral in~\eqref{eq:app_cont_solution}. With the common prior $\bm{P}_0^{-1}$,
\begin{equation}
\bm{J}_N \preceq \bm{J}(t_f)
\quad\Longleftrightarrow\quad
\bm{J}_N^{-1} \succeq \bm{J}^{-1}(t_f).
\label{eq:app_loewner}
\end{equation}
In particular, $\mathrm{tr}(\bm{J}_N^{-1}) \geq \mathrm{tr}(\bm{J}^{-1}(t_f))$: the discrete bound is the larger of the two. It reflects the information actually available from the finite measurement set, which is the quantity the trajectory optimization should minimize. Second, $\bm{g}$ is continuous along the trajectory, so the Riemann sum in~\eqref{eq:app_disc_solution} converges to the integral in~\eqref{eq:app_cont_solution} as $\Delta t \to 0$, and the two bounds coincide in the limit.

\subsection{Pareto-Based Weight Selection for Case N5}
\label{app:pareto}

This appendix demonstrates, on Case N5, the practical selection of the weights in~\eqref{eq:cost_functional} by a sweep of the single normalized weight $w_P$ in~\eqref{eq:cost_pareto}, in place of the manual per-case selection of $w_2$ reported in Sec.~\ref{sec:results}.

\begin{figure}[ht]
\centerline{\includegraphics[width=20pc]{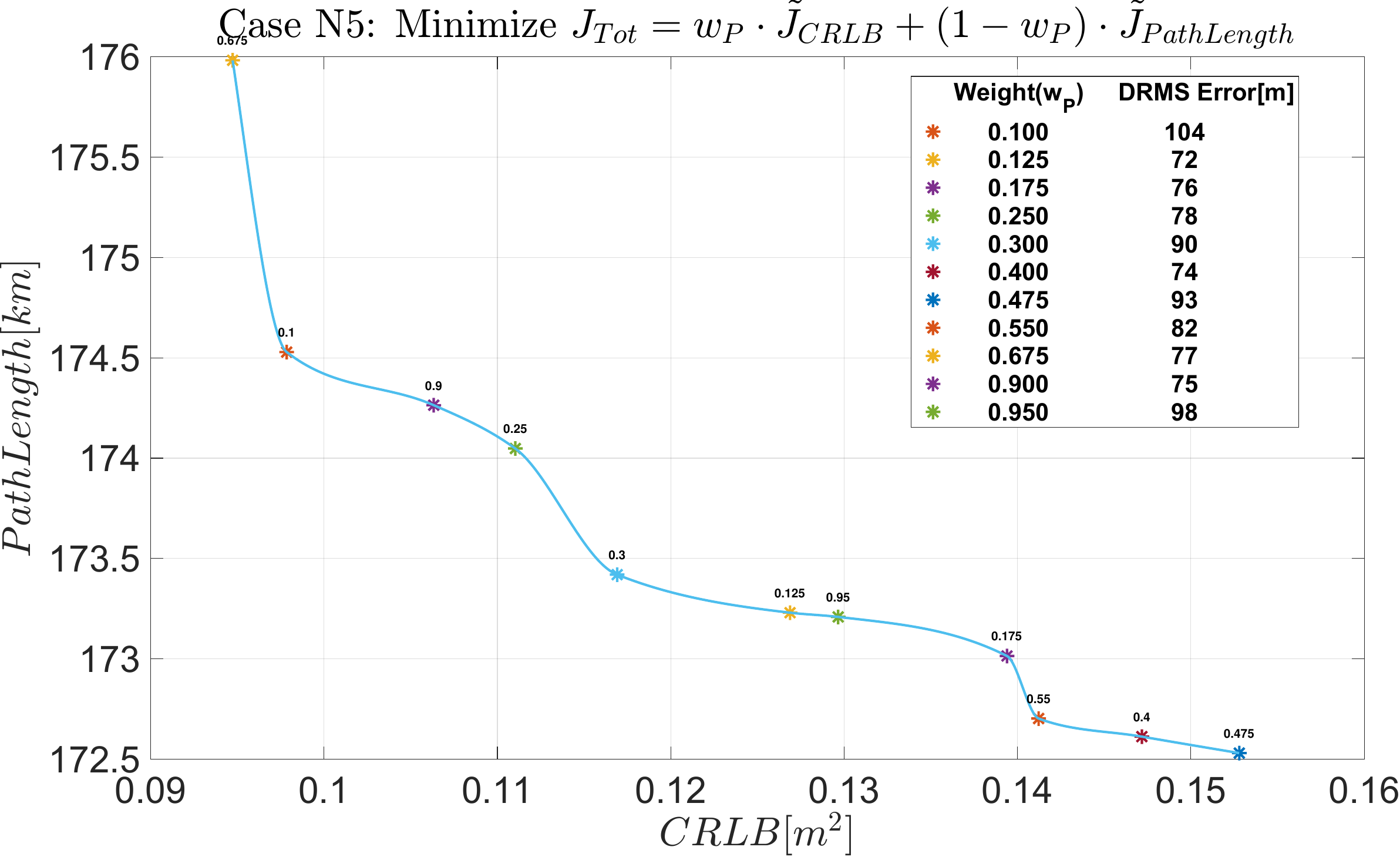}}
\caption{Case N5: non-dominated solutions of the sweep of $w_P$ in~\eqref{eq:cost_pareto}, in unscaled units. Point labels denote $w_P$; the legend lists the DRMS error of each solution from 200 Monte Carlo runs.}
\label{fig:pareto_N5}
\end{figure}

The scale factors were set to $S_1 = 1000$\,m$^{-2}$ and $S_2 = 0.001$\,m$^{-1}$. Two trial optimizations at $w_P = 0.5$ sufficed to select these values, chosen so that the two cost terms contribute comparable magnitudes at the trial solution~\cite{Ross2018}. With $S_1$ and $S_2$ fixed, the optimal control problem was solved for $w_P$ from 0 to 1 in steps of 0.025, a fixed budget of 41 solves, and the dominated solutions were discarded. The remaining non-dominated solutions are shown in Fig.~\ref{fig:pareto_N5} in unscaled units of path length and PCRLB trace, together with the DRMS error of each solution evaluated from 200 Monte Carlo EKF runs conducted under the simulation setup of Sec.~\ref{sec:results}. The $w_P$ values are not monotone along the front: the cost landscape is non-convex, DIDO returns the best solution found for each weight, and neighboring weights can converge to different local minima. The front is therefore an estimate assembled from the best solutions found, not an exact Pareto front.

Because the trajectory is planned offline from the map, the Monte Carlo evaluation is available at planning time, and the front point with the lowest DRMS error is selected: $w_P = 0.125$, attaining a PCRLB trace of 0.13\,m$^2$ and a DRMS error of 72\,m at a path length of 173\,km. The manually tuned solution of Table~\ref{tab:main_results} for the same case attains 0.15\,m$^2$ and 88\,m (1,000 Monte Carlo runs) at the same 173\,km path length. At equal path length, the swept solution reduces the bound by 13\% and the realized error by 18\%, and it replaces the per-case search over $w_2$ with a fixed sweep that carries over to the remaining cases unchanged.

\bibliographystyle{IEEEtaes}
\bibliography{bibliography}


\end{document}